\documentstyle[aps,pre]{revtex}           
\tighten                                  
\begin{document}                          
\draft                                    
\twocolumn                                

\title{Growth and Structure of Random Fibre Clusters and
Cluster Networks}

\author{N. Provatas$^{1}$, T. Ala--Nissila$^{1,3}$,
and M. J. Alava$^{2}$ }

\address{
$^1$University of Helsinki, Research Institute for Theoretical Physics,
\\P.O. Box 9, FIN--00014 University of Helsinki, Finland
}

\address{
$^2$  Laboratory of Physics,
Helsinki University of Technology, Otakaari 1M,
FIN--02150 Espoo, Finland
}

\address{
$^3$Brown University, Department of Physics, Box 1843,
Providence, R.I. 02912, U.S.A., and Tampere University of
Technology, Department of Physics, P.O. Box 692, FIN--33101 Tampere,
Finland
}

\date{September 3, 1995}

\maketitle
\narrowtext

\begin{abstract}

We study the properties of 2D fibre clusters and
networks formed by deposition processes. We first
examine the growth and scaling properties of single clusters.
We then consider a network of such clusters, whose spatial
distribution obeys some effective pair distribution function.
In particular, we derive an expression for the
two--point density autocorrelation
function of the network, which includes
the internal structure of a cluster and
the effective cluster--cluster pair distribution function.
This formula can be applied to obtain information about
nontrivial correlations in fibre networks.

\end{abstract}


There are many phenomena in nature that can be viewed
as deposition problems, brought about through
various transport mechanisms that bring particles to a surface.
They include
a multitude of processes such as
deposition of colloidal, polymer and fibre particles
\cite{Eva93,Pri95,Ono86,Dod94,Ala94,Nie95,Ryd92}.
Many deposition phenomena involve particles whose size is large
compared to the their mutual interaction range,
and so the main deposition mechanism is due to particle
exclusion.  Among the most studied
in this class is the Random Sequential
Adsorption model \cite{Eva93,Pri95}.
There particles are deposited on a surface
and either stick or
are rejected according to certain exclusion rules,
with a maximum coverage (the ``jamming limit'')
less than unity.
This is in contrast to multilayer growth \cite{Eva93,Nie95,Ryd92}.

Processes of particle deposition may often begin from
colloidal suspensions; solid particles suspended in a fluid.
For some such systems, the
interparticle repulsion is strong enough to prevent multilayer
growth \cite{Ono86}. However,
the existence of dispersion forces can cause the
particles to flocculate, or aggregate, and to precipitate out of
the suspension \cite{Ryd92,Mur95}. For larger particles or
clusters of particles, gravity often induces
sedimentation out of the suspension \cite{Sch95}.

A particularly interesting
class of deposition--related problems which has
received little attention
is the deposition of fibres, fibre clusters, and the formation of
fibre networks. Perhaps the most practical application
of fibre deposition is that of paper--making. During its formation
paper undergoes several stages, beginning as a colloidal suspension
of fibres and ending up as a deposition of fibres on a surface,
when the fluid is drained out.
There have been many attempts to model
the structure of fibre networks \cite{Dod94,Ala94,Hau74,Nor72,Kal60}.
Much of this work has focused on
the calculation of power spectra of ideal random fibre networks and their
subsequent comparison with mass distribution data obtained from
paper--making experiments.

In some sedimentation problems, such
as in the making of laboratory
paper sheets, the fibres
can flocculate while still in solution \cite{Ker92}.
This may occur for a variety of reasons, ranging from
mechanical fibre--fibre interactions to hydrodynamic
forces in the suspension \cite{Sch95}.
In such circumstances the resulting network may consist
of clusters of fibres. Similar clusters are also
expected to occur in other deposition problems.
This is one of the main motivations
for our work where
we present a detailed study of the statistical
properties of deposition processes involving individual
fibres and clusters.

We start with a simple model of deposition of
fibres of length $\lambda$ and width
$\omega$ on a 2D plane. Deposition begins
from an initial seed fibre. At each deposition
event only a fibre that overlaps at least one fibre
already in the cluster,
is kept. This growth rule is motivated
by the adhesive sticking of fibres.
The process forms a connected
cluster of $N$ fibres which is statistically spherically
symmetric. Thus it is possible to define
an average maximum radius $R(N)$.
Fig. \ref{clusterfig} shows one configuration of a computer generated
$N$--cluster on a lattice, where $N=2000$.
We find that for $N >> 1$ the radius satisfies

\begin{equation}
R(N) = B N^{\beta},
\label{Radius}
\end{equation}

\noindent where the exponent $\beta \approx 1/3$, over the range of $N$
examined and
for all fibre geometries simulated, while the constant $B$
depends only on
fibre dimensions.
Fig. \ref{radiusfig} shows a plot
of $R(N)$ for two different types of fibre geometries;
$\lambda \times \omega = 20 \times 1$
and $\lambda \times \omega = 50 \times 1$ showing the
corresponding plots on a log--log scale.
We note that Eq.~(\ref{Radius})
also gives the  number of fibres in a cluster directly from
the radius.

The exponent $1/3$ in Eq. (\ref{Radius})
also arises from the following simple argument.\footnote{This
result has been also independently derived
by J. {\AA}str\"om (J. {\AA}str\"om, Pro Gradu avhandling,
{\AA}bo Akademi (1989) (unpublished).}
Assume we have a cluster of average radius $R(N)$ composed of $N$ fibres
of linear dimension $a=\sqrt{\lambda \omega}$. Next if we
deposit an additional $\Delta N $ fibres onto the cluster
its radius grows by $\Delta R$ and area by $\Delta A$.
The change in the cluster area
$\Delta A = \frac{1}{2} a^2 \Delta N (2\pi Ra )/(\pi R^2) =
2 \pi R \Delta R$, which in the continuum limit
gives Eq.~(\ref{Radius}), with $B=(3/ 2 \pi)^{1/3}a$.
The factor $2 \pi Ra/\pi R^2$ comes from the ``active'' zone
at the edge of the cluster which is the only region where
added fibres increase the area.

Another measure to characterize the $N$--cluster is
the radial mass probability per unit length. This quantity
is given by
\begin{equation}
\rho_N (r) = \frac{m_N(r)}{ \int_0^{ \infty} m_N(r) dr },
\label{Mass}
\end{equation}
\noindent where $m_N(r)$ represents the mass per unit area at a radial
distance $r$ from the center of the cluster.
For large $N$,
the normalized
average mass probability density $P_N(r) \equiv \langle \rho_N (r)
\rangle$ can be written as
\begin{equation}
P_N(r)= \frac{1}{\tilde{\gamma} R(N)} f(r/R(N)),
\label{Denscale}
\end{equation}
\noindent where $\tilde{\gamma}
=\int_0^{\infty} f(u) du$ and where
the average implies an ensemble average over all
$N$--cluster configurations.
Figure \ref{densityfig}  shows a plot of $P_N (r)$ for five
different values
of $N$ while the scaling function $f(r)$ is
shown in the inset. The scaling function $f$ is
universal for the values of $\lambda$ and $\omega$ studied here.
The mass probability distributions have a peak--like component at the origin,
continuing almost linearly for the greater part of their width.

At the edge of the cluster, there is a fast decaying
region that also
scales as shown in Fig. \ref{densityfig}. This is
associated with the rough edge of the cluster which we can
examine
in the context of kinetic roughening of growing interfaces \cite{Bar95}.
Defining fluctuations of the radius as
$W(N)=\langle(\tilde{R}(N)- R(N))^2 \rangle^{1/2}$,
where $\tilde{R}(N)$ is the radius for one cluster,
we find it to grow as $W \sim N^{x}$, where $x$ is consistent with
1/3 as expected since $W/R=const.$
The roughness of the
cluster is expected to alter the percolation threshold of a random
fibre--cluster network from that of a distribution of
solid disks \cite{Lor93}.

We now consider a {\it disordered fibre network} constructed by
deposition of $N$--clusters. This model is completely general
and the only assumption is that there exists an
effective spatial pair distribution function
between any two clusters.
One plausible application of such a network
is the description of sedimentation of
fibre suspensions containing flocs
of fibres, as discussed above.

Starting from a distribution of $n_c$ fibre clusters,
labeled by an index $i$, and where the $i^{th}$ cluster
contains $N(i)$ fibres, the mass distribution of the resulting
network is $
M(\vec x) = \sum_{i=1}^{n_c} m_{N(i)} (\vec x - \vec x_i), \label{Cor1}$
where $m_{N(i)}$ is the mass density
and $\vec x_i$  the coordinate of the center of the $i^{th}$
cluster.
The two--point density autocorrelation function of
this network is given by \cite{Hau74}
\begin{equation}
G(\vec r) = \lim_{A \rightarrow \infty} F^{-1}
\left( \frac{  \langle |\hat{M}( \vec k )|^2 \rangle }{A}
\right)
\label{Cor2}
\end{equation}
\noindent where $A$ is the area over which the 2D network is deposited,
$F^{-1}$ is the inverse Fourier transform,
and the average is taken over
all network configurations.
The Fourier transform of $M(\vec x)$ is given by
\begin{equation}
|\hat{M}( \vec k )|^2 = \sum_{n=1}^{n_c} \sum_{m=1}^{n_c}
e^{ -2 \pi i \vec k \cdot (\vec x_n - \vec x_m) } I_{N(n)} I_{N(m)}^{*},
\label{Cor3}
\end{equation}

\noindent where $I_{N(n)}$ is the Fourier transform of
$m_{N(n)}(\vec x)$ evaluated over the area of the $N(n)$--cluster.
The coordinate
in $I_{N(n)}$ is defined relative to the center of the $n^{th}$ cluster.
Taking the ensemble average of Eq.~(\ref{Cor3}) we make the assumption that
the description of individual clusters is independent of the positions
of their centers. Thus
\begin{eqnarray}
\langle e^{ -2 \pi i \vec k \cdot (\vec x_n - \vec x_m) } I_{N(n)} I_{N(m)}^{*}
\rangle =
\hspace{3cm} \nonumber \\
\langle e^{ -2 \pi i \vec k \cdot (\vec x_n - \vec x_m) } \rangle \times
\langle I_{N(n)} I_{N(m)}^{*} \rangle.
\label{Cor5}
\end{eqnarray}

Next, making the assumption that internal
structure of any given cluster is independent of the others,
we write
$\langle I_{N(n)} I_{N(m)}^{*} \rangle = \langle I_{N(n)}
\rangle \langle I_{N(m)}^{*} \rangle$ noting
that $\langle I_{N(n)} \rangle$ implies an ensemble average over all
$m_{N(n)}(\vec x)$.
This average can be split into two parts: over all configurations
where the $n^{th}$ cluster contains $N(n)$ fibres and over
the number $N(n)$ itself.  Breaking up the average in this way
we define $\bar{m}_{\bar{N}} (r) \equiv \langle m_{N(n)} (\vec x) \rangle$
to be the average mass density of an
$\bar{N} \equiv \langle N(n) \rangle$ cluster.
Using Eq.~(\ref{Denscale})
and noting that the total average
mass in an $N$--cluster is $N w_f$, where $w_f$ is defined as the mass
per fibre,
we can show that
\begin{equation}
\bar{m}_{\bar{N}}(r) = \frac{w_f}{\pi \gamma B^3} R(\bar{N}) f(r/R(\bar{N})),
\label{Cor9}
\end{equation}

where $\gamma =2 \int_0^{\infty} u f(u) du$.
Thus we have
$\langle I_{N(n)} \rangle \langle I_{N(m)}^{*} \rangle
= | J_{ \bar{N} } (| \vec k |) |^2$,
where $J_{ \bar{N} } (| \vec k |) $ is the
Fourier transform of $\bar{m}_{\bar{N}}(r)$.
To evaluate the exponential term on the right hand side of
Eq.~(\ref{Cor5}) we consider two cases:
$n=m$ and $n \ne m$.  The first
gives unity, while the second depends on the type
of effective cluster--cluster interaction
between clusters $n$ and $m$.  Defining
$\Delta \vec{x}_{n,m} \equiv \vec x_n - \vec x_m$, we note that the ensemble
average over all cluster--network realizations is equivalent to integrating
$ e^{ -2 \pi i \Delta \vec x_{n,m} }$ over all $\Delta \vec{x}_{n,m}$, weighted
by a cluster--cluster pair probability density $g(\Delta \vec x_{n,m})$.
We also assume, by isotropy, that $g(\Delta \vec x_{n,m})$ is
independent of the indices $n$ and $m$, and radially symmetric \cite{Note2}.
The average of the exponential on the right hand side of
Eq.~(\ref{Cor5}) can thus be written as
\begin{eqnarray}
\langle e^{ -2 \pi i \vec k \cdot \vec \Delta x_{n,m}} \rangle =
\hspace{4cm} \nonumber \\
	\left\{\begin{array}{ll}
          \frac{1}{A} \int_A e^{-2 \pi i \vec k \cdot  \vec x}
	  g( |\vec x| ) d \vec x  & \mbox{$n \ne m$} \\
          1,                     & \mbox{$n=m$} \\
                  \end{array}
                   \right .
\label{Cor7}
\end{eqnarray}

Combining the equations above we finally obtain
\begin{eqnarray}
\langle |\hat{M}( \vec k )|^2 \rangle =  n_c | J_{\bar{N}} (k) |^2 +
\hspace{3cm}
\nonumber \\
\frac{n_c (n_c -1)}{2 A} | J_{\bar{N}} (k) |^2  \hat{g}(k),
\label{Cor8}
\end{eqnarray}

\noindent where $\hat{g}(k)$ is the Fourier transform of $g(|\vec x|)$.
Eq.~(\ref{Cor8}) comprises two
terms. The first represents the spectral density of
$n_c$ individual fibre clusters. Substituting this term into Eq.~(\ref{Cor2})
gives the two--point density correlation function of $n_c$ individual
$\bar{N}$--clusters.  The second term arises due to effective
cluster--cluster interactions. The
number of clusters $n_c$ can be expressed in terms of a parameter
$\eta$, defined as the number of fibres per unit
area \cite{note2}.
In particular, we have $n_c = \eta A / \bar{N}$.
Writing $\bar{N}$ in terms of the average cluster
radius  and substituting
Eq.~(\ref{Cor8})  into Eq.~(\ref{Cor2}),
the following $G_{\bar{N}}(r)$
is finally  obtained:
\begin{eqnarray}
G_{\bar{N}}(r)= \frac{4}{\gamma^2 B^3 \eta} R^3(\bar{N})
\int_0^{\infty} \Pi^2(k) k J_0(2 \pi k r) dk
\nonumber \\
+\frac{2}{\gamma^2} \int_0^{\infty} \Pi^2(k) \hat{g}(k) k J_0(2 \pi k r) dk,
\label{Cor12}
\end{eqnarray}
where $\Pi(k)$ (a dimensionless quantity) is given by
\begin{equation}
 \Pi(k) = \int_0^{\infty} f(z) z J_0(2 \pi k R(\bar{N}) z) dz
\label{Cor11}
\end{equation}
\noindent  and where
$J_0$ is the Bessel function of order zero.
In obtaining Eq.~(\ref{Cor12}) we normalized $G(r)$ by dividing
by the square of $\eta w_f$ \cite{Hau74,note2} and noted that
both $\Pi(k)$ and $\hat{g}(k)$ are radially symmetric.

Equation (\ref{Cor12}) is our main result for the fibre network.
The first component in it describes
the density--density correlations of a uniformly random distribution ($g(r)=0$)
 of
fibre clusters.  For this case it is easy to
show that Eq. (\ref{Cor12}) becomes $ G_{\bar{N}}(r) = R(\bar{N})
\Gamma (r/R(\bar{N}))$
where $\Gamma (r)$ is Gaussian for small $r$ and rapidly goes to zero as $r \gg
1$.
This dependence on $R(\bar{N})$ in
Eq. (\ref{Cor12}), allows us to extract the typical fibre--cluster size
directly from the correlation function.

The second term in Eq. (\ref{Cor12}) arises due to an effective
pair distribution function acting between clusters.
This function can lead to nontrivial long--range
correlations in the fibre network. Such correlations
could arise from hydrodynamics in the sedimentation
of particles from suspensions \cite{Sch95}.
We have examined a model
pair distribution given by
$\hat{g}(k) \propto k^{-\chi}$.  Figure \ref{intcorfig}
shows the second term of $G_{\bar{N}}(r)$ for
$\chi=1.1,1.3,1.6$ and $R(\bar{N}) =3$,
where distance is measured in units of $B$. For
$r > R(\bar{N})$ (where the
first term of $G_{\bar{N}}(r)$  becomes negligible), we find
$G_{\bar{N}}(r) \sim r^{-\alpha}$, where
the correlation exponent $\alpha <1 $ decreases continuously
as $\chi$ is increased from $0$ to $2$. Since the function $\Pi(k) \approx 0$
for
$k > 1/R(\bar{N})$, we expect any effective pair distribution satisfying
$g(k) \sim k^{-\chi}$ for small $k$, to give essentially
the same power law correlations at large $r$.

In conclusion, we have presented a model for the
statistical properties of disordered 2D fibre networks
formed by deposition processes.  The radius of fibre clusters
scales with the number of fibres.
Similarly, the density profiles of a cluster
obey scaling laws, with a universal scaling function for
the cases studied here.
Starting from the scaling form,
we have derived the two--point
density--density correlation function for a fibre network
due to deposition
of clusters, whose spatial distribution is given.
In the case of uniform, random deposition of fibre clusters
correlations exist over the range of the mean cluster size.
However, an effective cluster--cluster interaction
can give rise to long--range power law correlations. In this way
the final structure of deposited fiber networks may give us information
about the effective interactions between fibers in a suspension.

We expect that our $G_{\bar N}(r)$
can be compared with experiments on sedimentation
of fiber--like particles in suspensions. A prime example
of such a process is the formation of laboratory paper sheets,
whose final mass density can be measured
by the beta--radiogram technique \cite{Nor72}.
Our calculations will be compared with long--range
correlations observed in such radiogram
data in an upcoming publication.

Acknowledgements:   This work has in part been supported
by the Academy of Finland through the MATRA program.


FIGURES

\begin{figure}
\caption{A fibre cluster
made of 2000 fibres of dimension $\lambda \times \omega =50 \times 1$.
The cluster is constructed by depositing fibres randomly,
keeping only those that form one connected fibre cluster, beginning
from an initial fibre.}
\label{clusterfig}
\end{figure}

\begin{figure}
\caption{A plot of $R(N)$ on a log-log scale.
In the bottom curve $\lambda \times \omega =20 \times 1$,
while for the top curve $\lambda \times \omega =50 \times 1$.
The solid lines are drawn for comparison and have
a slope of $1/3$.}
\label{radiusfig}
\end{figure}

\begin{figure}
\caption{The radial mass probability per unit area,
for $N=1500, 2000, 2500, 3000$ and $4000$ fibres,
with $\lambda \times \omega = 50 \times 1$.  The inset shows the
scaling function $f(x)$ {\it vs.} $x=r/R$.}
\label{densityfig}
\end{figure}

\begin{figure}
\caption{The  correlation function of Eq. (10)
with $\hat{g}(k)= k^{-\chi}$, where $\chi=1.1,1.3,1.6$ from
bottom to top.  In the large $r$ limit power law behaviour
$r^{-\alpha}$, with $\alpha=0.92, 0.71, 0.4$, emerges.}
\label{intcorfig}
\end{figure}


\begin{thebibliography}{99}
\bibitem{note2} This quantity
is proportional to mass per unit area, known as the
basis weight in paper physics literature, see Ref. \cite{Dod94}.
\end{thebibliography}
\end{document}